\documentclass[10pt,letterpaper]{article}
\usepackage[top=0.85in,left=2.75in,footskip=0.75in]{geometry}

\usepackage{changepage}

\usepackage[utf8]{inputenc}

\usepackage{textcomp,marvosym}

\usepackage{fixltx2e}

\usepackage{amsmath,amssymb}

\usepackage{cite}

\usepackage{nameref,hyperref}

\usepackage[right]{lineno}

\usepackage{microtype}
\DisableLigatures[f]{encoding = *, family = * }

\usepackage{rotating}

\raggedright
\usepackage[aboveskip=1pt,labelfont=bf,labelsep=period,justification=raggedright,singlelinecheck=off]{caption}

\makeatletter
\renewcommand{\@biblabel}[1]{\quad#1.}
\makeatother

\date{}

\usepackage{lastpage,fancyhdr,graphicx}
\usepackage{epstopdf}
\fancyheadoffset[L]{2.25in}
\fancyfootoffset[L]{2.25in}

\begin{document}
\vspace*{0.35in}

\begin{flushleft}
{\Large
\textbf\newline{Green power grids: how energy from renewable sources affects networks and markets}
}
\newline
\\
Mario Mureddu\textsuperscript{1,2,3},
Guido Caldarelli\textsuperscript{2,4,5,6},
Alessandro Chessa\textsuperscript{2,6},
Antonio Scala\textsuperscript{4,2,5},
Alfonso Damiano\textsuperscript{7,2,*}
\\
\bigskip
\bf{1} Dipartimento di Fisica, Universit\`a di Cagliari, Italy
\\
\bf{2} IMT Institute for Advanced Studies, Lucca, Italy
\\
\bf{3} Department of Physics and Earth Sciences, Jacobs University, Bremen, Germany
\\
\bf{4} Istituto dei Sistemi Complessi (ISC), Roma, Italy
\\
\bf{5} London Institute for Mathematical Sciences, London, UK
\\
\bf{6} Linkalab, Cagliari, Italy
\\
\bf{7} Dipartimento di Ingegneria Elettrica ed Elettronica, Universit\`a di Cagliari,  Italy
\\
\bigskip

%
%

* alfio@diee.unica.it

\end{flushleft}
\section*{Abstract}
The increasing attention to environmental issues is forcing the implementation of novel energy models based on renewable sources. {This is } fundamentally changing the configuration of energy management and {is} introducing new {problems} that are only partly understood. 
In particular, renewable energies introduce fluctuations {which} cause an increased request {for} \emph{conventional} energy sources to balance  energy requests {at} short notice.
In order to develop an effective usage of low-carbon sources, such fluctuations must be understood and tamed.
In this paper we present a microscopic model for the description and  {for} the forecast of short time fluctuations related to renewable sources {in order to estimate}  their effects on the electricity market. 
To account for the inter-dependencies {in} the energy market and the physical power dispatch network, we use a statistical mechanics approach to sample stochastic perturbations {in} the power system and an agent based approach for the prediction of the market players' behavior. 
Our model is  data-driven; it builds on one-day-ahead real market transactions {in order} to train agents'  behaviour and allows {us} to {deduce} the market share of different energy sources. 
We {benchmarked} our approach on the Italian market, finding a good accordance with real data.


\section*{Introduction\label{sec:Intro}}

The increasing demand for energy, the improved sensitivity to environmental issues, and the need for a secure supply are all contributing to a new vision of energy resource  management\cite{allen2000reserve,david2000strategic}.
This new awareness is contributing to the development of a novel approach in energy planning, based on the rational use of local resources\cite{song2000optimal}.  In this contest, distributed energy management is considered one of the viable solutions to integrate local renewable sources and to promote rational  use of energy\cite{morales2010short}.
Moreover, the recent emphasis on sustainability, also related to  climate-change policies, {requires} a fast development in the use of renewable resources  in  local energy systems. This  determines a fast growth of distributed generation and co-generation, { there has not been a corresponding}  fast upgrade of the electricity infrastructure. This  inhomogeneous evolution of the different components of power systems is the consequence of the {present} structure of the electricity network, which, being characterised by strict dispatching and planning rules,  hardly fits with the increasing demand {for} flexibility connected to 
the distributed generation\cite{bouffard2008stochastic}. 

Such {a} scenario induces unavoidable effects on the electricity market which reveals an unexpected sensitivity to the enhancement of distributed generation based on Renewable Energy Sources 
(RES)\cite{kiviluoma2012short,li2012probability,shaaban2013dg}.
In fact, due to their non-programmable {characteristics} and  widespread geographic distribution, the development of RES-based distributed generation is undermining the technical and economic  models on which the electricity system are {currently} based. In particular, they highlight problems in the classical management model of energy-flows. In fact, the classical  hierarchical  and deterministic methodologies used to 
(i) manage the power system,  
(ii) forecast the energy demands and production,
(iii) balance the network, 
all show drawbacks which finally affect the electricity market price. 
An  example of such problems comes from the analysis of the effects {of policy of the subsidies}  granted by governments to promote the  exploitation of RES and for implementing the climate-change policies: in fact, such policies have played a major role in the {amplification} of critical market anomalies, like the negative and/or null price of electricity registered in the Germany and Italy. 
Therefore, to implement the new smart-grid paradigms, it is necessary to change and renew the classical approaches for modeling and managing the electricity market. 
However, even if the production of energy from renewable sources introduces perturbations in the power system and in the electricity market\cite{klobasa2010analysis}, it constitutes a crucial {advantage} in emissions trading.
Generators based on renewable sources have an {intrinsic high level of}  forecast uncertainty, highly variable both in time and space. Hence, the increasing amount of REnewable-like (RE) generators induce a stochastic variability in the system; 
such variability could induce security issues {such as} difficulties in voltage controlling\cite{ScalaLNCS2013} or unforeseen blackouts \cite{PahwaSciRep2014} and eventually causes a {significant} error in the power flow forecasting, {which} can give rise to extreme results {in the} energy market, {such as} very high prices or null/negative {selling price}.
To understand such effects, we must describe not only the dynamics of the fluctuations in energy production/demand but also the functioning of electricity markets.

{Current} electricity markets aim to reach efficient equilibrium prices at which 
both producers and distributors could sell electricity profitably.  
Typically, the electricity markets are hierarchically structured according to time-based criteria strictly connected to the power system's constrains.
In particular, the Short Term Market (ST) is usually structured into One-Day-Ahead Market (ODA), Intraday Market (ID) and Ancillary Service, Reserve and Balancing Market (ASR)\cite{joos2000potential}. 
Almost all the simulation approaches for the electricity market are either based on stochastic\cite{nogales2002forecasting,luh2003selecting} 
or on game theoretical\cite{song2002nash} studies based on past data-series, 
while few models focus on market equilibrium as obtained from production and transmission 
constraints\cite{gao2010electricity}. To the best of our knowledge,  nobody {has} yet addressed the effects of distributed generation, not only in power balancing but also in balancing market prices taking into account the network constraints. 

In this paper we present a simplified model that, taking into account the power system constraints, allows {us} both the forecasting of the balancing market and { the singling  out of the contribution} of various actors into the formation of the price.
Our model is data-driven, since information on the day-ahead market transactions is used to tune the agent-based simulation of the market behaviour. In our model we take as static constraints: (i) the grid topology; (ii) the type of production per node and (iii) the transmission rules. Our dynamical constraints are the maximum and minimum generation from power stations and their ramp variation (i.e. how fast  the amount of generated energy {can be changed}); such constraints influence factors like the short term availability of an energy source. 
The {inputs} of our model are typical consumer requests and the { forecast of geo-related} wind and solar energy generation.
We model the balancing market by introducing agents aiming to maximise their profits; such agents mimic the market operators of conventional generators\cite{bunn2001agent}. {The agents'} behaviour is modelled by a probability distribution for the possible sell/buy actions; such distribution is obtained by a training process {using} synthetic data.  
In each simulation, agents place bids on the balancing market based on the energy requests fixed by the ODA market. 
In order to ensure  system-security, the Transmission System Operator (TSO) selects the bids to guarantee energy balancing in real time. We model the TSO behaviour by choosing {bid} combinations according to the TSO's technical requirements and the economic merit order.
Our model allows {us} not only to forecast the statistics of the fluctuations in power offer/demand -- related to energy security -- but also the behaviour of the balancing market on a {detailed infrastructure knowledge  and to deduce} the market share of the various energy sources (e.g. oil, carbon etc.); hence, it has important practical implications, since it can be used as a tool and a benchmark for agencies and operators in the distribution markets.
Moreover, our modelling allows {us} to understand  changes in the market equilibria and behaviour due to the {increasing} penetration of distributed generation and {also} to address the question of economic sustainability of {specific} power plants.
As a case study, we present a detailed one-day analysis of the Italian electricity market.

\section*{Results\label{sec:Results}} 

{ RESs have a fundamental impact on  the functioning of the electricity market}  due to the technical constraints of  power systems, {which require} an instantaneous balance between  power production and demand. In fact, the electricity market is structured to {guarantee  matching between} the offers from generators and the  bids from consumers at each node of the power network according to an economic merit order\cite{allen2000reserve}.  
To perform this task, the exchanges starts one day ahead on the basis of  daily energy demand forecasting and then successive market sections refine the offers {with} the aim of   both satisfying the balancing {conditions} and of preserving the power quality and the security of energy supply. 

The most extensively studied market sector is the day-ahead market, which has been modelled both in terms of statistical analysis of historical data\cite{contreras2003arima}, game theory for the market phase\cite{song2002nash} and stochastic modelling of the market operators' behaviour\cite{nogales2002forecasting}.

On the other hand, few models have been proposed for the  last market section\cite{allen2000reserve,doherty2005newapproach,bouffard2006marketI,masohhour2011bidding},   devoted to {assuring} the \emph{real-time power reserve}. In fact, ASR allows {us} to compensate for the unpredictable events and/or the forecasting errors that can occur {to} the whole power system. In particular, the Balancing Market (BM) has a fundamental role in guaranteeing the reliability of the power system in presence of the deregulated electricity market.

The most important studies in ASR modelling aim to provide methods that allows the forecasting {of}  the amount of power needed for network stabilization purposes\cite{allen2000reserve,doherty2005newapproach,bouffard2006marketI,masohhour2011bidding},   whereas only {a} few research activities deal with the energy price forecast\cite{nogales2002forecasting}. 
With respect to the state of the art, we present in this paper an alternative model for daily BM time evolution. In particular, we reproduce the operator market strategies by means of an agent based approach, where agents represent typical market operators.

Our model is characterised by three phases: sampling of the perturbations, training of the agents, and {forecasting} of the balancing market.

In the first phase we use the information about the ODA and ID market to {deduce} realistic power flow configurations by taking into account the physical constraints of the {electricity} grid. We then introduce stochastic variations related to the {geographic} distribution of power consumption and RES generation; in such a way, we generate a statistical sample of configurations representing realistic and {geo-related} time patterns of {the} energy requests/productions to be balanced.
{In this paper we will consider only the simples stochastic model of variations, i.e. uncorrelated Gaussian fluctuations with zero mean and variances from historical data.} 
The difference at each time between the total actual power requests and the volume of the ODA+ID market is the size of the balancing market. 
The result of this phase samples the statistics of fluctuations induced by renewable sources; hence, it has important applications {for}  energy security (forecasting energy congestions and/or outages) and {for} maintaining quality of service.
We show in Fig. \ref{fig1} the setup of the system for the first phase; in particular, panel (a) shows the topology of the electricity transmission network in Italy, panel (b) shows the {division of the } Italian market  and panel (c) shows a typical daily time evolution of ODA+ID outcomes with the detailed contribution of each primary energy source.
{The} market zones are used for managing {possible} congestions occurring in the Italian electricity market.

In the second phase we use such balancing {requirements} together with the static and the dynamic constraints of conventional power {plants} to train the agents of the balancing market by optimising their bidding behaviour (see sec.\ref{sec:Methods}). 

{Here}  the balancing market generators are only conventional; hence, optimising the usage and implementation of renewable resources to diminish short time market fluctuations is crucial to augment the sustainability of power production.

In the third phase we use the balancing market size and the trained agents' biddings to evaluate market price evolution by performing a statistically significant number of simulations. In {these} simulations, each agent can place bids, both for positive (upward market) or negative (downward market) balancing needs.
This data is produced {throughout} the day at fixed intervals on a geo-referenced grid; for the Italian balancing market,  the bids are accepted each 15 minutes.
Typical {simulation} outputs in the upward and the downward electricity balancing market are:
\begin{itemize}
\item The time evolution of the balancing market size.
\item The time evolution of the electricity prices 
\item The market share for each technology type 
\end{itemize}

In fig. \ref{fig2} we compare the results of the model with real data of the actual upward  and downward balancing market obtained from the Italian market operator web site\cite{gmesite}; { the data reported in} \cite{gmesite} {are averaged} for each hour. We {have taken}  the 2011-2012 winter season {as a reference period}. In the upper panels of fig. \ref{fig2} we show that the predicted sizes of the  downward and upward markets -- expressed in term of energy reductions/increases for balancing requirement -- {match with the time-data series}  of the reference period. 
In the lower panels of fig. \ref{fig2} we show that the predicted prices in the  downward and upward markets also {match} with {the time-data series}  of the reference period. 
We notice that price and size have a similar shape, highlighting the expected correlation among sizes and prices.
To the best of our knowledge, this is the first time that {has been} possible to forecast the behaviour of the balancing market without using {a} historical time series analysis but using {information} coming out from the one-day ahead power system.

A {significant result}   of our approach is the forecast of the detailed contribution of each primary energy source to the downward  and upward electricity balancing market. 
In fig. \ref{fig3} we show that conventional energy sources contribute in a different manner to the upward and downward market. {For} example, due to {dynamic} constraints, carbon power plants' contribution is negligible (due to the limits in the minimum operative power generation, mostly in the upward market) even if their energy {production costs} are the lowest. This result highlights {the fact} that market shares {in the balancing market} do not depend only on energy costs but {stem} from an equilibrium {between dynamic response, energy costs, geographical position} and interactions among the different energy sources.

\section{Conclusions\label{sec:Discuss}} 
The use of renewable energy sources is creating a new energy market where it is of the utmost importance to be in  condition to anticipate trends and needs from users and producers to reduce inefficiencies in energy management and optimize  production.  
The future transformation of the traditional passive distribution network {into} a pro-active {one} is requiring the implementation of {an} energy system  where production and {power} fluctuations can be efficiently managed. In particular, {power} fluctuations have the strongest impact on markets and {on short-time} energy-continuity {requirements}.

Previous research on short time energy forecasting concentrates on next-day electricity prices, showing that the analysis of time-series yields accurate and efficient price-forecasting tools when using dynamic regression and transfer function models \cite{nogales2002forecasting} or ARIMA methodology \cite{contreras2003arima}. 
Systematic methods to calculate transition probabilities and rewards have also been developed to optimize market {clearing} strategies \cite{song2000optimal}; to improve market clearing price prediction, it is possible to employ neural networks \cite{luh2003selecting}.
A further step toward and integrated model of (day ahead) market and energy flows has been taken in \cite{bouffard2006marketI}, where authors propose a market-clearing formulation with stochastic security assessed under various conditions on line flow limits, availability of spinning reserve and generator ramping limits.
However, one-day ahead markets and balancing markets are fundamentally different and need separate formulations \cite{allen2000reserve}.

Since wind power is possibly the most erratic renewable source, it has been the focus of most investigations {into} short-time fluctuations. The analysis of possible evolutions in optimal short-term wind energy balancing highlights the needs {for} managing reserves through changes in market scheduling {(higher and more regular)} and in introducing stochastic planning {methods} as opposed to deterministic   ones\cite{kiviluoma2012short}. In \cite{Vrakopoulou2013probabilistic}, together with a probabilistic framework for secure day-ahead dispatch of wind-power, a real-time reserve strategy is proposed as a corrective control action. On the {operator's} side, the question regarding the virtual power plant (i.e. a set of energy sources aggregated and managed by a single operator as a coherent single source)  participation {in} energy and spinning reserve markets with  bidding strategies that {take} into account distributed resources and network constraints {have} been developed in\cite{masohhour2011bidding} {utilizing complex}  computational solutions like nonlinear mixed-integer programming with inter-temporal constraints solved by genetic algorithms.

In this paper we model both the electric energy flows and the very {short-term} market size taking into account the variability of renewable {energy} generation and customer demands {through} a stochastic approach. Network and ramping constraints are explicitly taken into account {through} the AC power-flow model while market price prediction is modelled through an agent-based simulation of energy operators. The inputs of the model are the  day-ahead prices and sizes, quantities that {are} possible to successfully predict \cite{nogales2002forecasting,contreras2003arima}. Our approach falls {into} the class of models of inter-dependent critical infrastructures \cite{BookNON2014}. We validate our model in the case of the Italian power grid and balancing market; we {found} that even a simplified stochastic model of production and demand based on uncorrelated Gaussian fluctuations allows {us} to predict the statistics of energy unbalances and market prices.
Our model complements the virtual-plant approaches that concentrate on the marketing strategies of single operators managing several sources.
To the best of our knowledge, the explicit mechanism through which fluctuations enter  the price determination {had} never been considered explicitly before our investigation.

In the {current} phase of transition from a centralised to a distributed generation system,  our approach allows {us} to address the complex task of estimating  the additional cost associated {with} the balancing of renewable energy sources. {This} evaluation  {allows us } to better understand the real impact of green sources in diminishing the carbon footprint, since balancing -- in absence of a well-developed technology of energy storage -- still relies heavily  on conventional generators. Moreover, by comparing the current situations with novel scenarios where new generators (nodes in the model) are introduced, our approach allows for a detailed geo-localised  \textit{\``what-if\''}  analysis of the energy planning. An important direction {for our model  to develop}   would be a deeper understanding and modelling of fluctuations. In fact, the probability-distribution of fluctuations in energy production has different statistics {depending on} the renewable type. Moreover, both spatial and temporal correlations {among the fluctuations} should be taken into account: as an example, weather-influenced fluctuations -- like the ones from wind and solar generators -- display naturally a cross correlation among nearby located sources; on the same pacing, the non-instantaneous character of weather variations also induces temporal correlations. Though our analysis stems from a theoretical approach to understanding the effect of stochastic components in an interconnected system, it has immediate practical implications since the computational burden of our method is compatible with the scheduling time of the balancing  market, permitting the potential use of this software for \textit{\``on-the-fly\''} decision support.

An important development {for} our model would be to address \``\textit{what-if}\'' analysis aimed {at understanding} how the introduction of new rules and policies affects the market. In fact, it has been shown that regulatory intervention {affects -- using cash-out arrangements} -- not only spot price dynamics, but also price volatility (i.e. fluctuations) \cite{Facchini2014}. Moreover, by predicting {power umbalances}, our approach allows for a better understanding of the energy security risks induced by renewable sources. In fact, the introduction of the stochastic components is crucial for the management of electrical energy systems, for which the deterministic approach {has allowed a detailed description of the functioning of the electrical energy system}, by virtue of (i) an accurate profile for the management of generation and (ii) a high degree of accuracy in load-prediction, i.e. conditions that are nowadays {significantly} changed.

\section{Methods\label{sec:Methods}}

The development of models that allow the evaluation of ancillary service {costs} in an electricity system during a RES-based transition phase, has practical implications, particularly important in  energy system planning.
Moreover, the associated tools can be {usefully} implemented by the TSOs and the market operators {in order to} forecast in real time both the expected amount of energy required for balancing purposes and their  price evolution.

In previous studies\cite{contreras2003arima,nogales2002forecasting}, market sizes and the electricity price forecasts have been evaluated by statistical analysis of {time-data series. Despite  the  accuracy} of these methods, their formulations {did} not allow the forecasting of the possible changes in markets caused by a transformation of the system involving market rules and/or infrastructure {evolutions}  (different power grids topology, transmission codes, new or different management of power plants).
{We} propose a methodology that is able to take into account any upgrade, since it models the behaviour of the market operators subject to a realistic set of perturbations of the {current} system. 
The reference configuration of the power system is obtained starting from two datasets. 
The first dataset is related to the characteristics of the power system (from the TERNA website\cite{terna}) and includes the geo-referenced position of every 220 and 380 kV substations  with their electrical characteristics, the geo-referenced position of conventional generators  with their power rates and power ramp limits and the electrical characteristics of the power network.
The second dataset (from the GME website\cite{gmesite}) reports the detailed time evolution of  production/consumption for each 15 minutes of a reference day in the winter period 2011-2012.

Since we {aimed} to describe the entire electricity balancing market session, we {performed} a complete simulation for each of the market subsections. 
For each interval of 15 minutes, the simulation is characterised by three phases: sampling of the perturbations, training of the agents and {forecasting} of the balancing market.

In the first phase, the electric state of {the} power grid is initially  perturbed stochastically by adding uncorrelated zero-mean Gaussian fluctuations in order to mimic the variability both in the  power production  and in the demand. In particular, we employ realistic values of the variances of RES generators and of electricity demand to obtain a realistic set of perturbed physical states of electricity network. To each of these perturbed states, it corresponds an unbalanced power condition. The first phase allows {for} the statistical forecasting of the size of the balancing market.

In the second phase, the forecast sizes of the balancing market are used {in order} to train the agents to tune their offer propensities, i.e. their willingness to offer a certain amount of energy at a certain price. These propensities are, in {practice, due to the} expertise of the operators in understanding market fluctuations and placing  {bids that enable} them to reach  the maximum profit. To a further details on the operators' behaviour model, please refer to the supplementary informations. 

In the third phase, trained agents place bids on a set of realistic perturbations representing the possible balancing requirements; price is formed according to TSO's merit order. 

The implementation  of the proposed methodology requires a detailed description and analysis of the power system from a technical and economical point of view. In particular, the evaluation of the perturbed states in a medium-size national power transmission grid is a complex task; in our Italian case-study, it {involved} around a thousand  interconnected nodes dispatching power, around a hundred  conventional generators, around a thousand  RES generators, and  thousands of loads.
Each node {was subjected} to complex physical constraints which {had} to be modelled adequately in order to ensure a correct description of the system. In addition, global system constraints {had to} be considered in order to ensure the correct behaviour of the system in {terms} of the quality of the supply.
Moreover, the distributed RES generators and loads which {were} aggregated at the corresponding transmission nodes, {assumed} values of power that {fluctuated} in time and space. We {modelled} their power production or consumption in a statistical way, assuming Gaussian-like forecast errors with standard deviations $\sigma_i$, which {represented} the expected power variations at each single node $i$ of the power grid in a given time. The application of {an} AC power-flow algorithm\cite{MatPower} {allowed for} the validation of the dynamic and static physical {constraints},  giving  the possible states of the power system.  
The considered variables associated {with} them are:
\begin{itemize}
\item Load power demand $D_l$ and the corresponding $\sigma _l$;
\item Wind power production $G_w$ and the corresponding $\sigma _w$;
\item Photovoltaic power production $G_{PV}$ and the corresponding $\sigma _{PV}$. 
\end{itemize}
The system variability {was} tackled {though} a statistical mechanics approach; the set of possible states of the system at a defined time {was} numerically sampled by adding {a random value extracted from a Gaussian distribution with zero mean and variance $\sigma _i$} to the expected power production and consumption at every node $i$  and RES generator of the grid.
Each perturbed state {was} characterised by a different total power production $G^{tot}$ and demand $D^{tot}$, and their difference $S = G^{tot} - D^{tot}$ {was} the required balancing power; hence, $S$ {was} a random variable that {represented} the market size. To sample the statistics of the market behaviour, we {needed} a significant number of possible balancing requirements. {With this aim, we generated} 6000 statistically independent perturbed states for each time interval since such number of sampling is enough to get a sufficient accuracy in our statistics.

In order to model the balancing market, the specific rules on which it is based  should be described. In general, a market session is an auction, in which the bids placed by market operators are accepted by the TSO according to a cost-minimization method. The operative rules of the Italian balancing market are briefly described in the supplementary information section.
For each perturbed state, an auction {is} made with a corresponding sampled value $S$ of the market size. Since $S$ can be either positive or negative we can have respectively a so-called upward market or a downward market session.

In a market session, each agent (market operator) $k$ represents a conventional power plant and can place a bid $(p_k, g_k)$ {at} the auction, in which it specifies the amount of energy $g_k$ that the corresponding power plant can provide to the system, and its price $p_k$. Once the bids have been placed, the TSO accepts all the viable offers until the total energy needed for balancing is {reached}.

Since the bid values are obtained {related} to the agent $k$ propensities described by a specific probability distribution $M_k$, agents must be trained to estimate their propensities. To {this end, we started} from an initial guess $M_k=M^0$ and {performed} several market sessions in which each agent {updated} its propensities in order to maximise a profit function as described in detail in the supplementary {information} section. 

Once the agents have been trained, we can forecast the behaviour of the balancing market by performing market sessions on the sampled perturbed states. In addition to the market size $S$, we can calculate the global price per kilowatt $P=\sum p_k / S$ from the set of accepted offers. Notice that $P$ is a random variable, associating a market price to each perturbed state of the system.                          

The outputs of the simulations are the sampled distributions $\mathbf{P}^{up}$, $\mathbf{P}^{down}$, $\mathbf{S}^{up}$ and $\mathbf{S}^{down}$ of sizes and prices of the upward and downward markets. Fig. \ref{fig4} shows a flow-chart of the whole simulation procedure.
In order to describe the system evolution {over time, these} distributions have been obtained for each time interval $t$, obtaining a dynamic distribution of market size and energy price. 
In order to validate the outcomes of our simulations with the available balancing market outcomes\cite{gmesite}, results {have} been aggregated for each hour of the day.

\section*{Acknowledgments\label{sec:acknowledgments}}
Any opinion, findings and conclusions or reccomendations expressed in this material are those of the author(s) and do not necessary reflect the views of the funding parties.


\paragraph{Competing financial information}
The authors declare no competing financial interests.

\begin{figure}[h]
\includegraphics[width=1\textwidth]{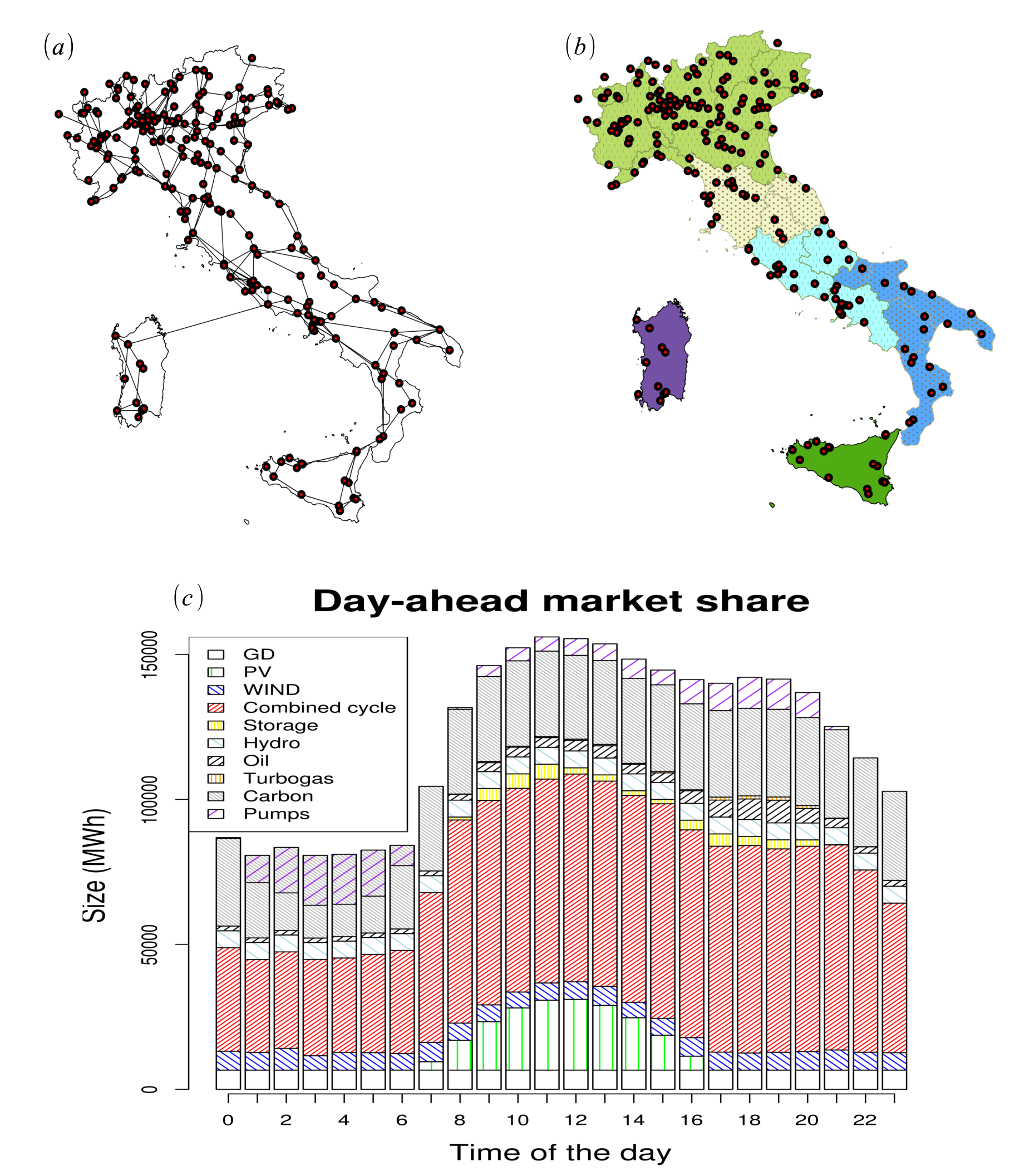} 
\caption{
{\bf Input elements for evaluating the size of the balancing market (i.e. the short time fluctuations in power generation/demand). }
In panel (a) we show the electric transmission network in Italy; the topology and the physical characteristics of lines and generators are the constraints that influence the power flow. 
In panel (b) we show the market zone splitting used for managing the congestions of the entire Italian network.
In panel (c) we show a typical ODA+ID market (One-Day-Ahead Market + Intraday Market) final output with the  detailed contribution of each primary energy source; hence, it represent the day ahead energy needs as foreseen from energy operators. The balancing market takes care of short time fluctuations that occur during the day respect the scheduled ODA+ID output.
\label{fig1}}
\end{figure}

\begin{figure}[h]
\includegraphics[width=1\textwidth]{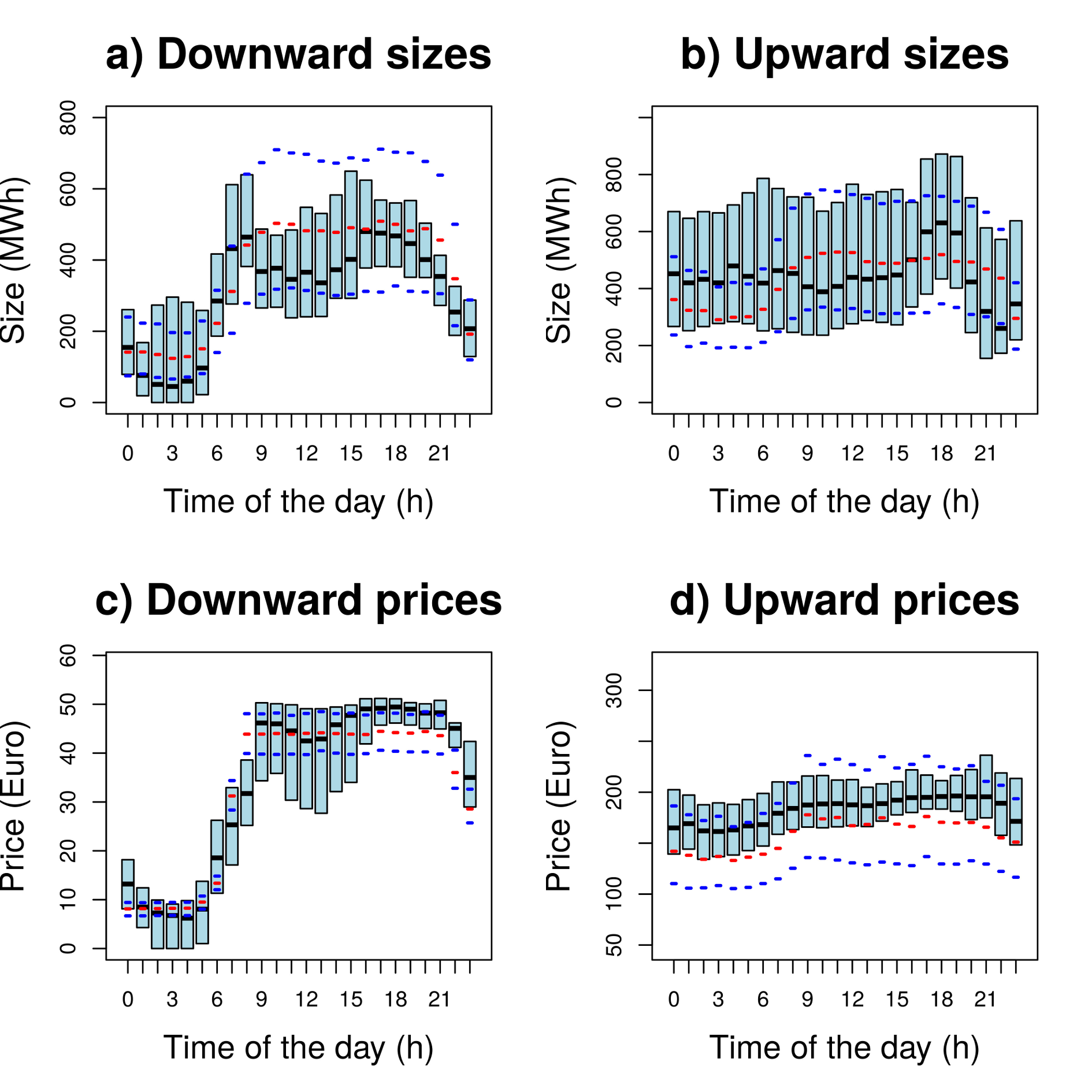} 
\caption{
{\bf Comparison among the results of our model and the real balancing market.} 
Upper left panel (a): Sizes of the downward market. 
Upper right panel (b): Sizes in the upwnward market. 
Bottom left panel (c): Prices of the downward market. 
Bottom right panel (d): Prices in the upwnward market. 
Notice that unbalanced power can either lead to (i) the necessity of producing \emph{less} power than what foreseen (left panels, downward markets) or to (ii) the need of more power than what foreseen (right panels, upward markets). 
Full rectangles represent the $1^{st}-3^{rd}$ quartile range (i.e. data is inside such range with $50\%$ probability) of the real data; the black segment in the full rectangles is the median of the real data.
Red segments correspond to the median and blue segments define the range from the $1^{st}$  to the $3^{rd}$ quartile of the data synthetically generated from our models.
In the upper panels we show the comparisons among real data and the predicted size of upward and downward markets, i.e. the difference among the foreseen energy production and the actual request.
In the lower panels we show the comparison among real market prices and the ones predicted from our agent based model.
\label{fig2}}
\end{figure}

\begin{figure}[h]
\includegraphics[width=1\textwidth]{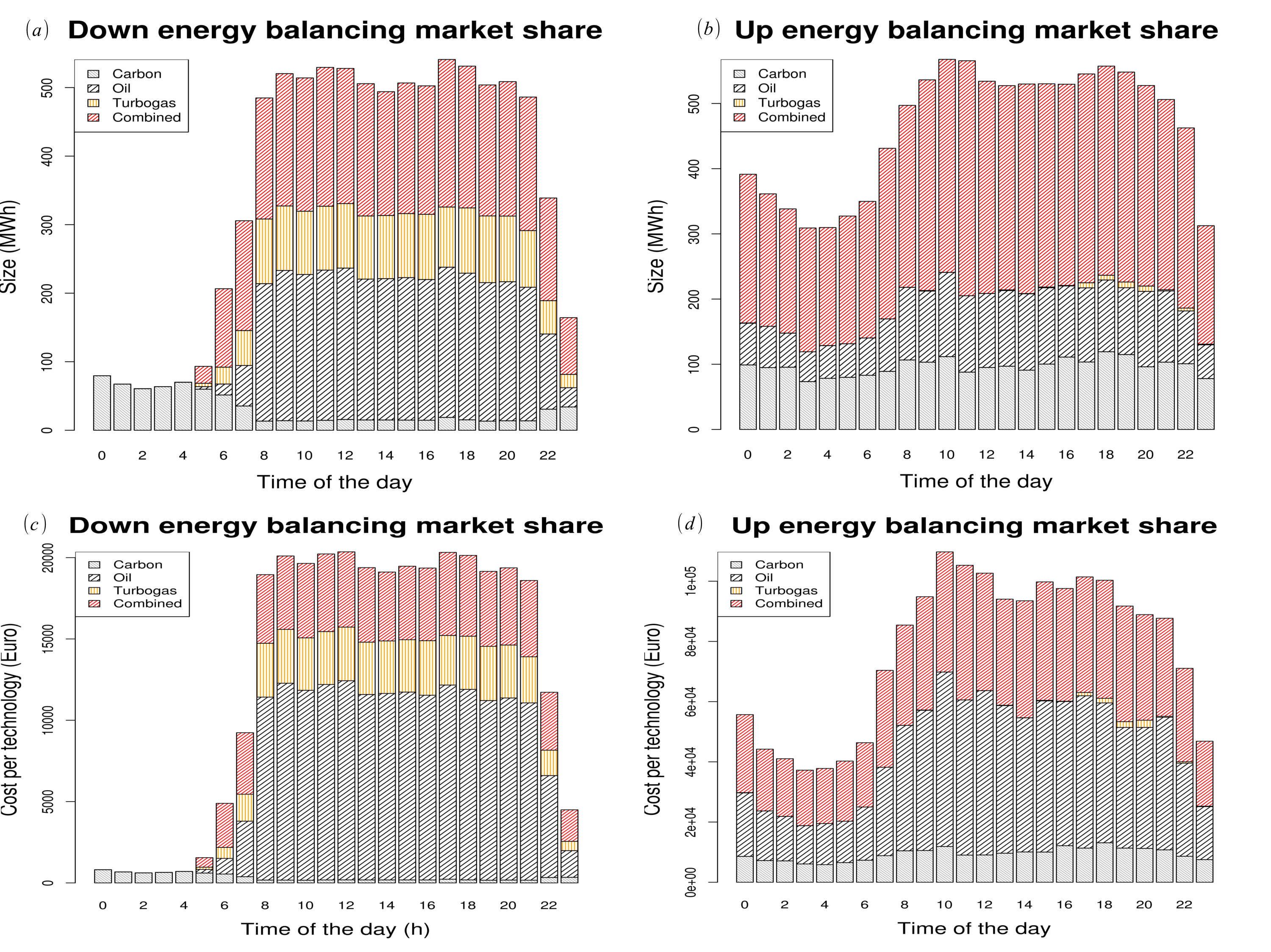} 
\caption{
{\bf Estimated total energy (upper left panel (a) downward market, upper right panel (b) upward market) and total cost (bottom left panel (c) downward market, bottom right panel (d) upward market) in the balancing market;} notice that our model is able to detail the contribution of each conventional power plant technology. As expected, due to low ramping (i.e. slowness in changing operational conditions), carbon sources have a very low impact on the balancing market even if they have often the lowest costs.  
\label{fig3}}
\end{figure}

\begin{figure}[h]
\includegraphics[width=0.7\textwidth]{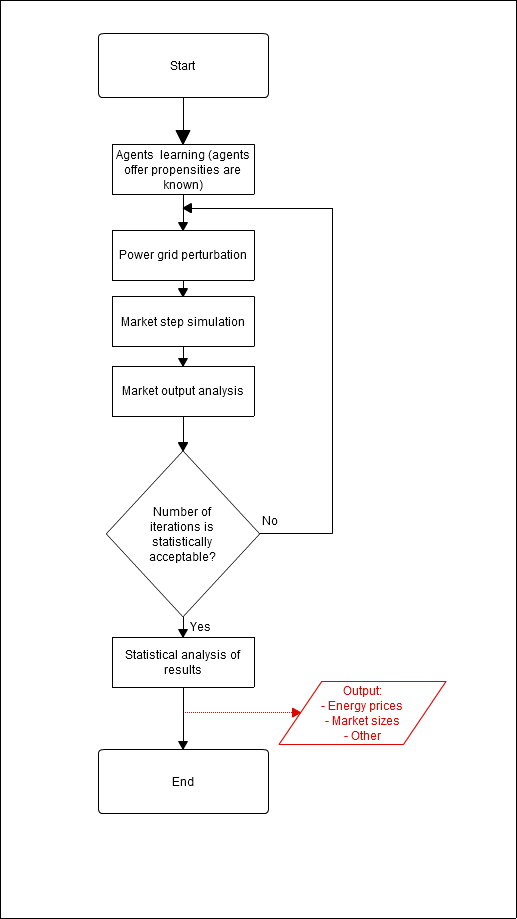} 
\caption{
{\bf Coarse-grained flow-chart of the simulation procedure for the coupled electric network - electricity market model.}
\label{fig4}}
\end{figure}

\section*{Appendix}


\subsection*{The Italian balancing market}
The Italian electricity balancing market is a pay-as-bid like market. In such markets, the goods are payed for the amount of money agreed in the placed bid.
The BM stage follows the one day-ahead market (ODA) and  intra-day market (IM). In ODA and IM the Market Authority (GME, Electrical Market Authority) decides the amount of energy that each generator must provide to the network, given economic and physical constraints and the forecast power production and consumption of loads and RES generators.

Since there is an the uncertainty associated to power production and consumption, after the ODA and IM market stages there will be a real time market stage, the balancing market (BM), whose target is to balance unpredicted changes in production (or load) that can occur respect to power production and consumption forecasts.
In the BM stage, the transmission system operators (TSO) 
interacts with authorized market operators, i.e. electrical operators (or group of them) that can offer or buy energy. These operators are often represented by a broker, whose job is to submit a bid on the market, specifying the quantity and the price of the energy that the operator plans to sell (or buy) on the BM at each time of the day. After the bidding stage, the TSO communicates to the brokers if their offers are accepted and how much power must be supplied to the system. Normally, there are various market sessions every day (in Italy there are 5 sessions per day, each related to offers spaced by 15 minutes intervals).
Power balancing is carried out on a zonal basis. There are 6 zones defined on a geographical basis, as shown on Fig. \ref{fig1}(b).

The Italian electricity balancing market is organized in 2 phases:
\begin{enumerate}
\item Auction: each broker places an upward and a downward bid for every time interval $t_n$ related to the auction. Every bid is a couple of real numbers $(p_{off}, g_{off})$, in which the broker specifies the amount of energy $g_{off}$ that the corresponding operator can supply to the system and the associated  price $p_{off}$. When a bid is accepted in the upward market, the operator agrees to provide the energy $g_{off}$ to the system by increasing his production by the same amount; on the downward market, the operator will instead reduce his production by the agreed amount $g_{off}$.
\item Market: Given the balancing power needs for every time interval $t_n$, the TSO accepts offers until the system power balance is reached, seeking economic advantage and with the condition that electrical constraints are met.
\end{enumerate}

\begin{figure}[h]
\includegraphics[width=1\textwidth]{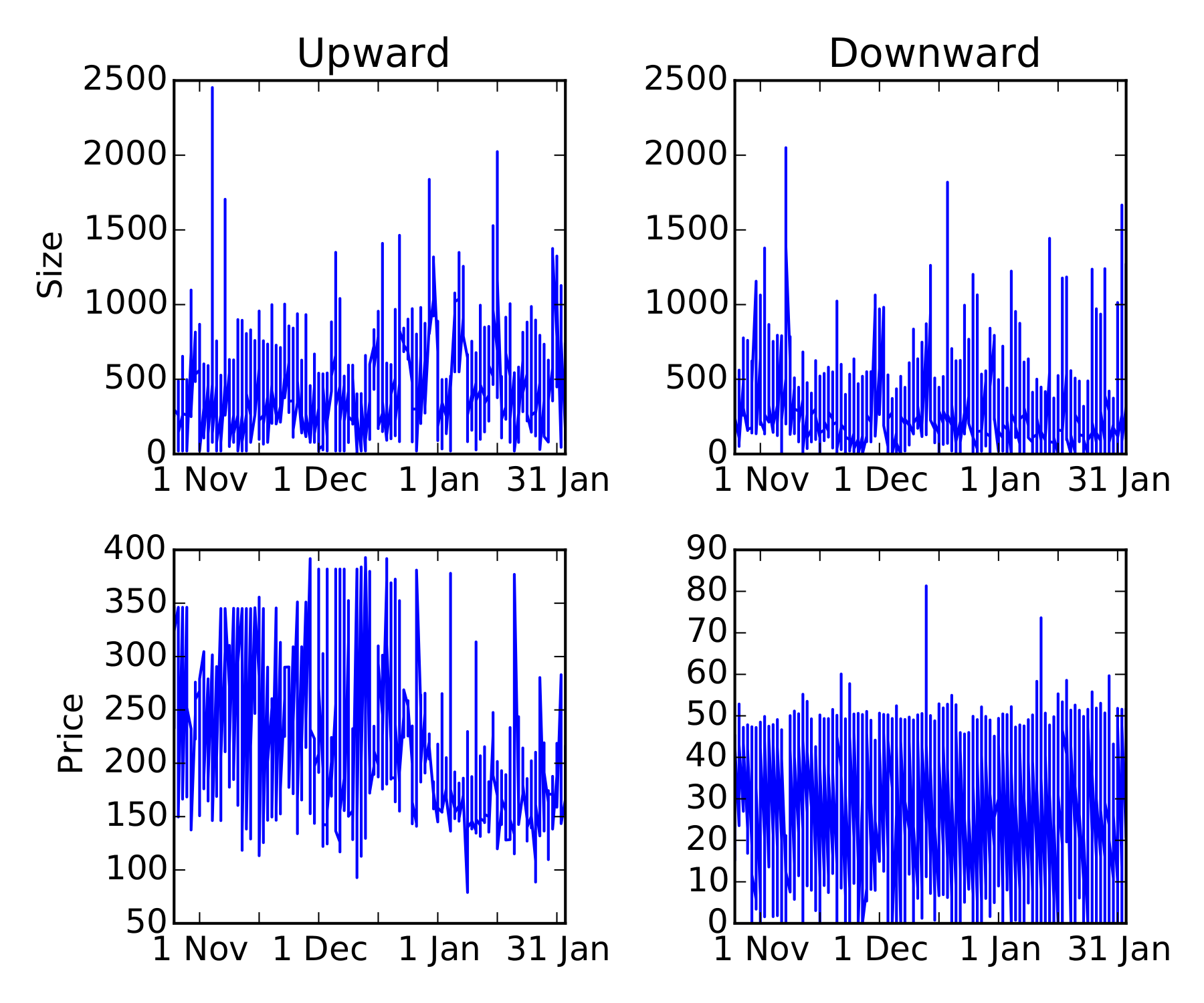} 
\caption{
{\bf Daily values of aggregated market sizes and prices for the reference months.} 
\label{figS1}}
\end{figure}

In fig.\ref{figS1} 
we show the time series for the aggregated Italian market sizes and prices.

\subsection*{Agent based simulation of electricity balancing market}

We base our agent model on the Roth-Erev algorithm\cite{pentapalli2008comparative}; such kind of algorithms have already been applied for simulating the Italian ODA electricity market\cite{Rastegar2009}. In such kind of models, agents learn how to place optimal bids in competitive auctions with the aim of buying (or selling) in the most convenient way. 

The behaviour of real operators is related to their market knowledge, often obtained by a learning process performed during time. Roth-Erev algorithms simulate this learning process by adjusting propensities using a self-consistent methodology whose goal is to  maximize profits. In this paper we apply a modified version of Roth-Erev algorithm as introduced by Nicolaisen et al\cite{Nicolaisen2001}. Since we don't have the information on the exact relationships among market operators and brokers, we consider every conventional power plant generator as a single agent. 

We describe operator propensities using a statistical description of the possible bidding strategies. The bidding strategies of the operator $k$ are described by a finite discrete set $\mathbf{S_k} = \{(m^i_k, s^i_k)\}$.  Here $0<i<N$ is the strategy index, $N$ is the number of possible strategies, $s^i$ is the operator propensity to offer at a given markup value $m^i_k$  ($1\leq m^i_k \leq 10$ for upward bids, $0\leq m^i_k \leq 1$ for downward bids); in our simulations $N=50$. The mark-up value allows to calculate the bidding price as  $p_{off} = C_{prod}\cdot m^i$, where $C_{prod}$ is the production cost (per MWh) of each generator, given by his technology type. The behaviour of the operators is modelled by a stochastic process in which the probability of placing a bid at a given price $p_{off} = C_{prod}\cdot m^i$ is the normalised propensity $q^i = s^i/\sum_i s^i$. 

An agent $k$ can offer an amount of power $g^k_{off}$ that must meet the following constraints:
\begin{itemize}
\item $G^k_{min} \leq G^k_{given} + g^k_{off} \leq G^k_{max}$: every generator has a minimum $G^k_{min}$ and a maximum $G^k_{max}$ of allowed power supply; $G^k_{given}$ is the actual power production of the generator.
\item $-G^k_{ramp} \leq g^k_{off} \leq G^k_{ramp}$: due to construction and technological limits, each generator has ramping constraints  that limits in time their maximum change in power production $G_{ramp}$. 
\end{itemize}

\begin{figure}[h]
\includegraphics[width=1\textwidth]{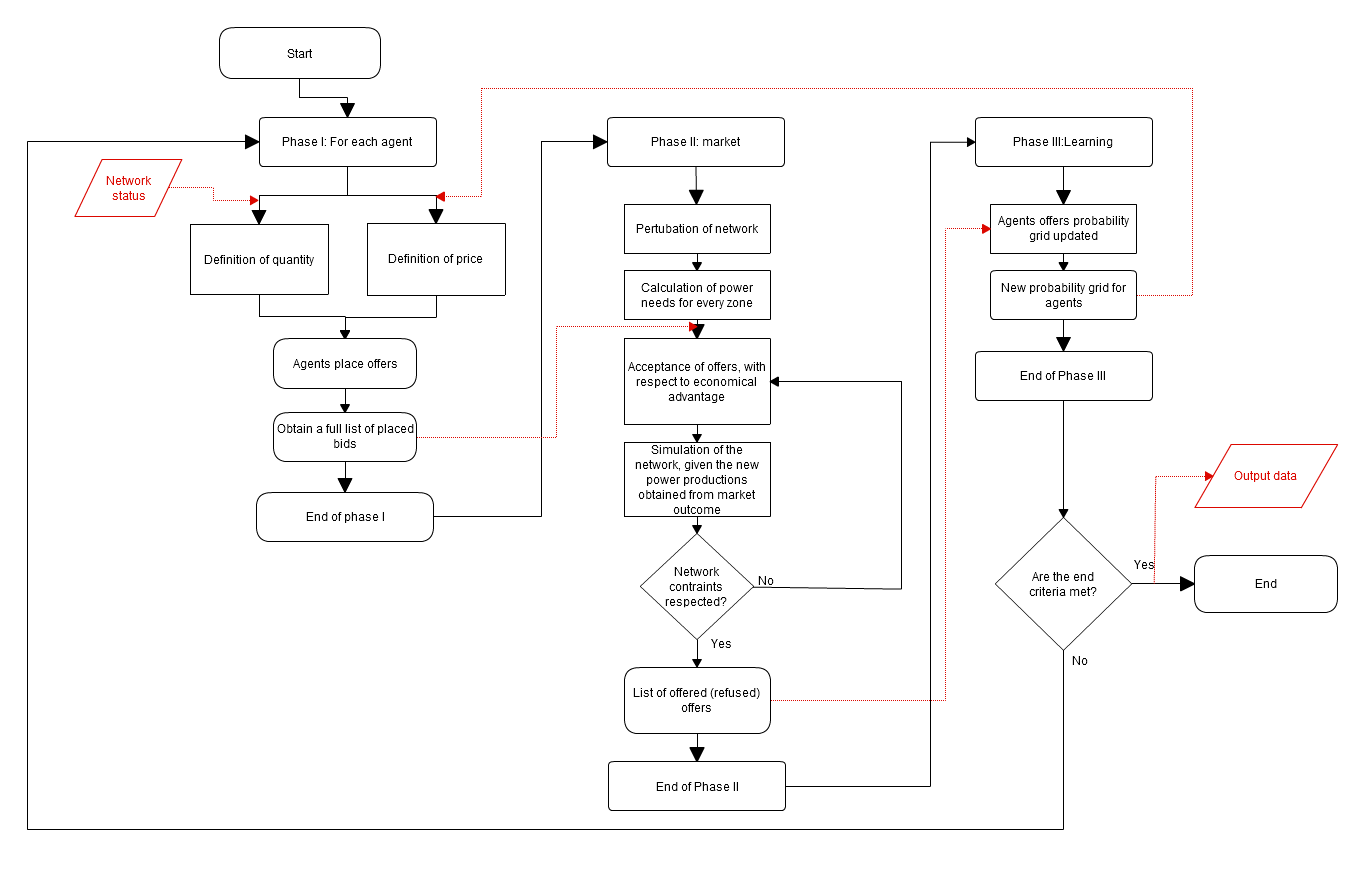} 
\caption{
{\bf Coarse-grained flow-chart of the simulation procedure for the market phase.} Notice that the couplings among the electric network and the electricity market must be taken into account during the simulation.
\label{figS2}}
\end{figure}

To optimise the propensities of the agents, we apply an iterative algorithm.  At the beginning of the learning algorithm, all propensities $s_n$ have the same value $s_n=1$. The iterations of the algorithm are divided in three phases:

\begin{enumerate}
\item Bid presentation: Every agent presents a bid $\left(g_{off},p_{off}\right)$, both for upward and downward market. This bid is given by a feasible quantity of offered energy $g_{off}$ (i.e. satisfying the physical constraints) and by a price $p_{off}$ that will be drawn from agents' propensities.
\item Market session: Given the knowledge of the balancing needs of the system, the TSO accepts all the bids needed to ensure that energy while seeking economic profit, verifying that the physical constraints of the system are met.
\item Agent update: Market outcomes are communicated to each agent, that updates his propensities in relation to the profit made in the session. Agents propensities at iteration $t$ are updated as follows:
\begin{equation}
s_i(t)=(1-r)\cdot s_i(t-1) + E_i(t)
\end{equation}
where $r \in [0,1]$ is a memory parameter and $E_i(t)$ is obtained from the relation:
\begin{equation}
E_i(t)= \begin{cases} 
(p_{off}-C_{prod}) \cdot g_{off} & \mbox{if bid has been accepted at time }t \\ 
e \cdot m_i(t-1)/\left(N-1\right)  &  \mbox{otherwhise}
\end{cases}
\end{equation}
where $e \in [0,1]$ is an experimental parameter that assign a different weight to played and non-played actions. 
\end{enumerate}
Fig.\ref{figS2} 
shows a flow-chart of the market model implementation.

To the best of our knowledge, Roth-Erev algorithms have been always applied by training agents over historical data. In this paper we overcome the need of historical data by training the agents on realistic system states that are synthetically generated.

\nolinenumbers



\end{document}